\documentclass[12pt,a4paper]{article} 
\pdfoutput=1
\usepackage{jheppub}
\usepackage{enumerate}
\usepackage{epsfig} 
\usepackage{float} 
\usepackage[caption = false]{subfig}
\usepackage{wasysym} 
\usepackage{mathrsfs} 
\usepackage{amsfonts} 
\usepackage{amsbsy} 
\usepackage{amscd} 
\usepackage{pstricks} 
\usepackage{multirow} 
\usepackage{tikz}
\usepackage{color}
\usepackage{slashed}
\usepackage{array}
\usepackage{tablefootnote}

\usetikzlibrary{arrows,positioning,shapes.geometric} 
\usepackage{color}
\definecolor{myred}{rgb}{0.6,0,0} 
\definecolor{myblue}{rgb}{0,0.2,0.4}
\definecolor{mygreen}{rgb}{0,0.9,0.1}
\definecolor{hc}{rgb}{.9,0.1,0.7}
\definecolor{hcout}{rgb}{.9,0.7,0.9}
\definecolor{Orange}{rgb}{1.,0.65,0.}

\numberwithin{equation}{section}
\numberwithin{figure}{section}
\numberwithin{table}{section}

\makeatletter
\newcommand{\fmslash}[2][0mu]{%
  \mathchoice
    {\fmsl@sh\displaystyle{#1}{#2}}%
    {\fmsl@sh\textstyle{#1}{#2}}%
    {\fmsl@sh\scriptstyle{#1}{#2}}%
    {\fmsl@sh\scriptscriptstyle{#1}{#2}}}
\newcommand{\fmsl@sh}[3]{%
  \m@th\ooalign{$\hfil#1\mkern#2/\hfil$\crcr$#1#3$}}
\makeatother

\newcommand{\lsim}{{\;\raise0.3ex\hbox{$<$\kern-0.75em\raise-1.1ex\hbox{$\sim$}}\;}}
\newcommand{\gsim}{{\;\raise0.3ex\hbox{$>$\kern-0.75em\raise-1.1ex\hbox{$\sim$}}\;}}
\newcommand{\met}{{\fmslash E_T}}

\usepackage{array}
\newcolumntype{C}[1]{>{\centering\arraybackslash$}p{#1}<{$}}
\usepackage{bm} 
\newcommand{\be}{\begin{equation}}
\newcommand{\ee}{\end{equation}}
\newcommand{\bes}{\begin{equation*}}
\newcommand{\ees}{\end{equation*}}
\newcommand{\bea}{\begin{eqnarray}}
\newcommand{\eea}{\end{eqnarray}}
\newcommand{\beas}{\begin{eqnarray*}}
\newcommand{\eeas}{\end{eqnarray*}}
\newcommand{\dmtotal}{\Delta M_{\tilde{t}\chi_1^0}}
\newcommand{\dmone}{\Delta M_{\tilde{t}\chi_1^+}}
 \newcommand{\dmtwo}{\Delta M_{\chi_1^+\tilde{\ell}}}
\title{Constraining slepton and chargino through compressed top squark search}
\author[a]{Partha Konar,}
\author[b]{Tanmoy Mondal,} 
\author[a,c]{and Abhaya Kumar Swain} 
\affiliation[a]{Theoretical Physics Group, Physical Research Laboratory, Ahmedabad - 380 009, India}
\affiliation[b]{Regional Centre for Accelerator-based Particle Physics,
Harish-Chandra Research Institute, HBNI,
Chhatnag Road, Jhunsi, Allahabad - 211 019, India} 
\affiliation[c]{Department of Theoretical Physics, Indian Association for the Cultivation of Science, 2A \& 2B Raja S.C.
Mullick Road, Kolkata - 700 032, India}
\emailAdd{konar@prl.res.in}
\emailAdd{tanmoymondal@hri.res.in}
\emailAdd{abhaya@prl.res.in}
  
\abstract{
We examine the compressed mass spectrum with sub-TeV top squark ($\tilde t$) as lightest colored (s)particle in natural supersymmetry (SUSY). Such spectra are searched along with an additional hard jet, not only to boost the soft decay particles, also to yield enough missing transverse momentum. Several interesting kinematic variables are proposed to improve the probe performance for this difficult region, where we concentrate on relatively clean dileptonic channel of the top squark decaying into lightest neutralino ($\chi_1^0$), which is also the lightest supersymmetric particle. 
In this work, we investigate the merit of these kinematic variables, sensitive to compressed mass region extending the search by introducing additional states, chargino and slepton (sneutrino) having masses in between the $\tilde t$ and $\chi_1^0$.
Enhanced production and lack of branching suppression are capable of providing a strong limit on chargino and slepton/sneutrino mass along with top squark mass. We perform a detailed collider analysis using simplified SUSY spectrum and found that with the present LHC data $M_{\tilde t}$ can be excluded up to 710 GeV right away for $M_{\chi_1^0}$ of 640 GeV for a particular mass gap between different states.
}
\preprint{HRI-RECAPP-2017-012}

\begin{document}
\maketitle
\newpage

\section{Introduction}

The Standard model (SM) of particle physics has been successfully validated in numerous experimental observations with great precision. The latest being the discovery of the SM Higgs boson by 
CMS~\cite{Chatrchyan:2012ufa} and ATLAS~\cite{:2012gk} at the Large Hadron Collider (LHC).
Another important objective of a high energy and high luminosity collider like LHC is to search for 
new physics and till now LHC has not provided any clinching evidence of new physics. With a strong perception that the SM is  a low energy effective theory, lots of effort is poured into constructing models which not only explain experimental observations like dark matter, neutrino oscillation etc, also accommodate some theoretical prejudices like naturalness. 
Supersymmetry (SUSY) undoubtedly is one of the most explored model in this regard.
In a R$-$parity conserving (RPC) SUSY, the lightest SUSY particle (LSP) is absolutely stable and is a natural DM candidate. 
In the RPC scenario typical SUSY particles are expected to produce in pair at collider experiment. After cascade decay depending upon possible interactions as well as mass hierarchy, result into a pair of LSPs which are undetectable at the detector. Consequently, SUSY signatures are mostly associated with production of large missing transverse momenta.

The non-observation of any signature of new physics at the LHC effectively constrains a big chunk of SUSY parameter space. Exclusion limits are rather prominent in case of conventional colored particles like squarks and gluino.  Currently, light top squark has been excluded up to 1 TeV 
for the neutralino mass below 160 GeV~\cite{ATLAS-CONF-2017-020,CMS-PAS-SUS-16-050}. But in order to stabilize the Higgs boson mass from large quantum correction,  at 
least one of the top quark partner (top squark) should be less than one TeV and expected to 
show up in the current run of the LHC. Incidentally, these strong constraints from the LHC on top squark weaken considerably if one considers a compressed mass spectrum having narrow mass gap between different sparticle states. One can analyse the situation with full supersymmetry spectrum squeezed enough satisfying available constraints from Higgs measurement and dark matter~\cite{Dutta:2015exw, Dutta:2017jpe}. However, conventional searches are done considering only simplified two level system  where the top squark, the next to lightest supersymmetric particle (NLSP) has mass difference in different degree with the LSP. This region of parameter space is extremely difficult to search at the LHC because of the indistinct kinematics from the SM backgrounds which makes it very challenging to disentangle them. A brief review of the search regions and different strategies pursued in the past is discussed below.
The top squark search regions are categorized depending upon the decay kinematics decided by the mass 
difference between the top  squark and LSP. There are mainly two search regions as follows:

\textit{Uncompressed top squark}:- In this region, the top squark ($\tilde{t}$) is heavier compared to the  lightest neutralino ($\chi_1^{0}$) and top mass  
combined, $M_{\tilde{t}} > M_t + M_{\chi_1^{0}}$ which allows the two body decay, 
$\tilde{t} \rightarrow t \chi_1^{0}$ or $\tilde{t} \rightarrow b \chi_1^{+}$~\cite{CMS:2017vbf} depending on the model. The dominant backgrounds come from top pair, associated top and weak boson pair production. These channels are probed widely by employing  numerous kinematical variables, for example, the transverse mass variable, $M_{T2}$~\cite{Lester:1999tx,Cho:2007qv,Bai:2012gs,Kilic:2012kw,Cao:2012rz}, the partonic Mandelstam variable, $\sqrt{\hat{s}_{min}}$~\cite{Konar:2008ei} and its variants~\cite{Konar:2010ma,Aad:2012yr}, scalar sum of the transverse momentum of all the reconstructed objects, $H_T$~\cite{Chatrchyan:2013wxa}, missing transverse momenta (MET), $\met$, effective mass $M_{eff}$~\cite{Hinchliffe:1996iu}, invariant mass of various visible particles~\cite{Hinchliffe:1996iu,Gjelsten:2004ki}, Razor variable~\cite{Rogan:2010kb,Chatrchyan:2012uea,Khachatryan:2015pwa}, $\alpha$ variable~\cite{Randall:2008rw} and its variant $\alpha_T$~\cite{Khachatryan:2016pxa} etc. Since the top quark in the final state can be very energetic primarily because of the large mass gap between top squark and neutralino, the top quark is boosted and as a result its decay products can be collimated. Jet substructure techniques and large radius fat jet methods were also studied~\cite{CMS:2017vbf}.
	
 \textit{Compressed top squark:-} This is the region of parameter space in which the top squark and neutralino mass difference is less than or equal to top quark mass, $M_{\tilde{t}} \lesssim M_{t} + M_{\chi_1^{0}}$. Since the mass difference between the top squark and the neutralino is small, the visible decay products produced through off-shell decay are very soft and often fail the selection criteria.  Therefore, the principal search channel at the LHC is missing transverse momenta along with a hard jet(ISR)~\cite{Drees:2012dd,Carena:2008mj,Bornhauser:2010mw,Delgado:2012eu,Krizka:2012ah,Ferretti:2015dea}. If the top squark has considerable lifetime, it can form stoponium bound state which eventually decays to SM particles. Searching for the stoponium~\cite{Drees:1993yr,Martin:2008sv,Duan:2017zar} is shown to be sensitive to the compressed region of top squark. Compressed top squark is also searched for the case in which it is produced in association with a $Z$ boson or a Higgs from the decay of the heavier top squark assuming that is accessible at the collider~\cite{Perelstein:2007nx,Ghosh:2013qga,Khachatryan:2014doa,Kang:2017rfw}. Compressed top squarks are also explored along with intermediate sleptons and charginos leading to multi-lepton final state~\cite{Padley:2015uma,Cheng:2017dxe}.

 This compressed region can further be classified into two separate categories depending on the mass difference between the top squark and neutralino. If the mass gap is greater than or equal to $W$ boson and $b$ quark mass combined, $M_{\tilde{t}} - M_{\chi_1^{0}} > M_W + M_b$, then the top squark decays through three body off-shell top decay with on-shell $W$ boson. If the mass gap is even smaller than $W$ boson mass, $M_{\tilde{t}} - M_{\chi_1^{0}} < M_W$, the top squark decays via four body decay where $W$ is also off-shell, $\tilde{t} \rightarrow b f f^{'} {\chi_1^{0}}$ or through loop induced flavor violating two body decay, $\tilde{t} \rightarrow c {\chi_1^{0}}$~\cite{Hikasa:1987db, Boehm:1999tr, Muhlleitner:2011ww}. Since the charm tagging is not efficient for the flavor violating two body decay, both the CMS and ATLAS rely on the monojet plus missing energy search~\cite{Khachatryan:2015wza, Khachatryan:2016dvc, Aad:2014nra, Aad:2015pfx}. Corresponding limit on top squark mass is 550 GeV using 13 TeV 36 $\textrm{fb}^{-1}$ data as reported by the CMS~\cite{Sirunyan:2017wif}.
 Similarly, in the leptonic decay channel in the region of the four body decay of the top squark, $\tilde{t}\to\,b\,\ell\,\nu_\ell\,\chi_1^{0}$, the CMS and ATLAS collaborations have performed several analyses which includes single lepton and dilepton search in addition to moderate missing transverse momentum~\cite{Khachatryan:2015pot, Aad:2014kra, CMS-PAS-SUS-16-031, CMS-PAS-SUS-16-025}. The current limit  for the single lepton search~\cite{CMS-PAS-SUS-16-052} on the top squark mass is about 500 GeV with 95$\%$ CLs for a mass difference of 30 GeV between the top squark and the LSP. For the dilepton search the limit is 400 GeV with the mass difference 40 GeV~\cite{ATLAS-CONF-2017-034}. All these limits are calculated with integrated luminosity of  36 $\textrm{fb}^{-1}$.

Recently the variables $R_M$~\cite{An:2015uwa} was proposed which is well suited to the compressed region where $M_{\tilde{t}} \simeq M_{t} + M_{\chi_1^{0}}$. It exploits the kinematics of the compressed region to explore the hadronic decay of pair of top squarks which recoil against a hard ISR jet. $R_M$ is defined as the ratio between missing transverse energy and transverse momentum of the hard ISR jet, $\met/P_T^{ISR}$. In a compressed region, top squark and the neutralino moves 
in equal velocity~\cite{Hagiwara:2013tva} and thus the variable essentially peaks at the ratio between the neutralino and top squark mass. Background from the top pair production does not follow  this kinematics and falls exponentially from zero with respect to $R_M$.
Subsequently, this idea is generalized to accommodate either semi-leptonic or di-leptonic decays of the top squark~\cite{Cheng:2016mcw, Cheng:2017dxe}. In general,  the 
presence of additional source of missing energy due to presence of neutrinos deteriorate the behavior of the $R_M$ leading to  decrease in its efficiency. For the semi-leptonic decay, the 
neutrino contribution can be subtracted using the available kinematic constraints in each event. Whereas  di-leptonic decay channel remains under constraint and thus providing a range of values, but its extremum values found to be useful~\cite{Cheng:2017dxe}.
Lately, two new variables $R_{bE}$ and $R_{\ell E}$~\cite{Konar:2016ata} were proposed for the compressed region of top squark decaying dileptonically. The search 
strategy using these observables proved to be quite useful in giving a complimentary limit to experimental bounds. Although they were advocated for the four body 
decay of top squark, these variables are not limited to a particular decay channel as well as a region of parameter space, details about these variables are 
discussed in the section below. In spite of all these efforts, the current limit at the LHC for this region is still weak especially for the leptonic decay channel 
of the top squark.

In this paper, we investigate the dilepton channel of the compressed top squark. While this signature is relatively clean and reliable in presence of light leptons (electron, muon), small leptonic branching ratio in this channel constraints weakly compared to the hadronic processes. Hence the limits coming from typical search with a simplified model where top squark as NLSP is still quite low despite different interesting kinematic variable. Strikingly different result is extracted if one considers two new states chargino and the slepton (sneutrino) in between the top squark and lightest neutralino LSP still within the compressed region. In fact, additional two states make the mass hierarchy even more squeezed. Recently proposed new variables, $R_{bE}$ and $R_{\ell E}$ are found to be extremely effective in this highly compressed scenario. With these observables, the top squark search limit can be significantly improved even for the current LHC data. In addition, these limits can also be translated to constrain the chargino and slepton (sneutrino) masses and we got bounds where there are no or extremely weak experimental limit exist in the parameter space.

The rest of the paper is organized in following way. In section~\ref{framework} we discuss the phenomenological model and the basis for different benchmark points we consider for present analysis. Motivating from distinct signal and background topology suitable kinematic variables are also constructed. Numerical set up, simulation for both the signal and background processes are described in section~\ref{simulation}. We also present the effectiveness of different kinematic variables considering representative benchmark scenarios. In section~\ref{results}, we present our main results in terms of statistical significance for all the benchmark points we consider. Experimental reach in terms of different masses and mass gaps are demonstrated. Finally, in section~\ref{summary} we conclude with final points.

\section{Model and Framework}\label{framework} 

%

\begin{figure}
 \centering
 \includegraphics[keepaspectratio=true,scale=0.6]{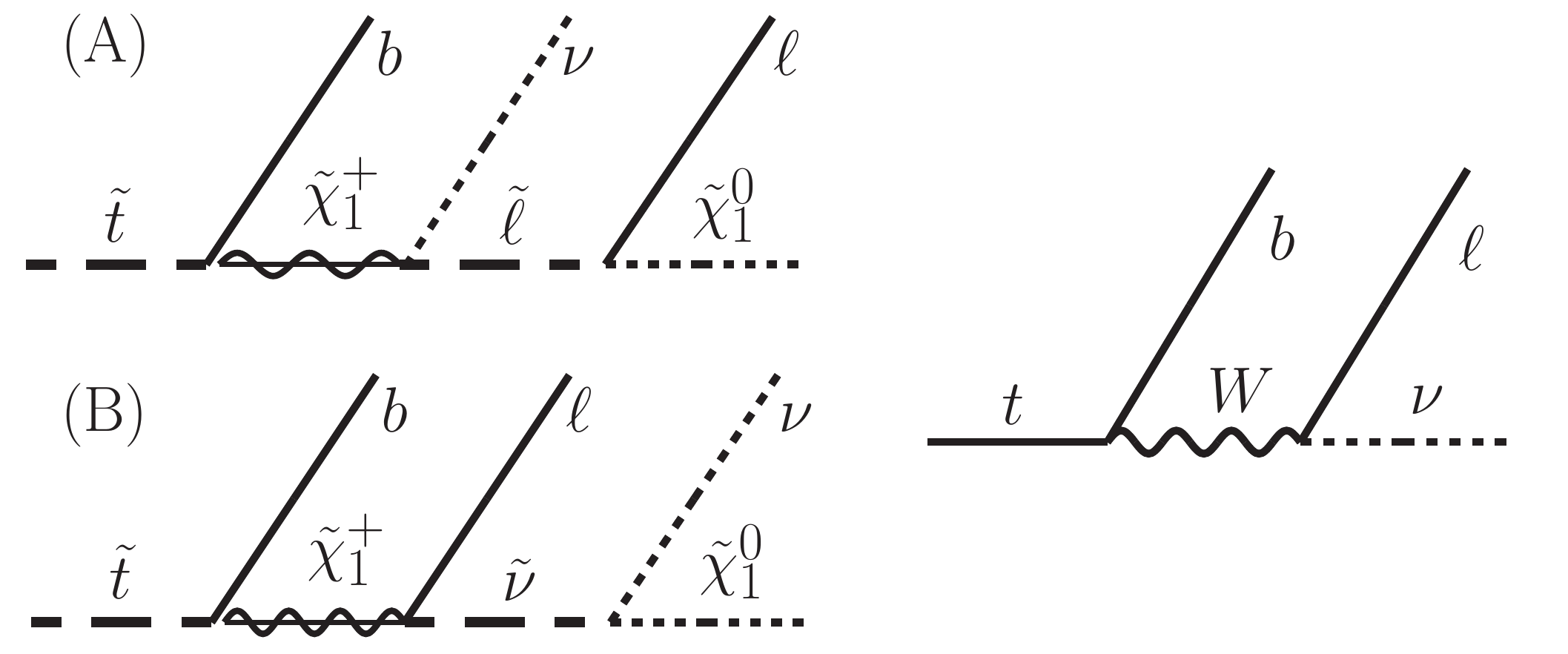}
 \caption{Left panel depicts decay toplogy of the top squark into dileptonic final 
 states via chargino and slepton and sneutrino mediator. For simplicity we show decay 
 of one of the pair produced top squark. In right panel we show decay of the dominant 
 background process $t\bar{t}$. }
\label{fig:decay-chain}
\end{figure}

In this analysis, we consider a scenario where the top squark decays predominantly through  the dileptonic decay mode.	
This can be realized, for a SUSY spectrum, if the first two 
generations of sleptons/sneutrino ($\tilde{\ell}$ / $\tilde{\nu}$) and chargino ($\chi_1^\pm$) lie between the lightest colored (s)particle (LCP), $\tilde{t}$ and the LSP ($\chi_1^0$).
Although there are various ways to realize the simplified model, 
one simple possibility can be a pMSSM scenario where the lightest neutralino is mostly 
bino type and the lightest chargino is wino type. In this case correct relic abundance 
can be achieved via bino-slepton coannihilation if the mass gap between neutralino and 
slepton is small ($<$ 15 GeV). For higher mass gap region, one can use the A-funnel region 
with $m_A\sim 2 m_{\chi_1^0}$ to achieve correct relic density. The Higgs mass constraint can also be 
satisfied easily by appropriately choosing the trilinear coupling and $\tan\beta$. We have assumed 
left selectron and left smuons are light and other states are heavier than the lightest stop. 
We fixed gaugino mass parameter $M_1$ and changed $M_2$ along with other soft SUSY breaking masses to 
achieve required mass gaps between $\chi_1^0$ and $\chi_1^\pm$ and lightest stop. 
The representative diagrams for the dileptonic decay modes of the top squark are shown in the left side of figure~\ref{fig:decay-chain}.
The top squark initially decays to $b$-quark and chargino, and subsequently the chargino decays to either slepton in association with a neutrino or a sneutrino along with a lepton. Finally, the slepton (sneutrino) decays to neutralino and a lepton(neutrino). To ensure that the top squark decays predominantly to dileptonic channel we assume that the mass differences respect the suitable mass hierarchy, $M_{\tilde{t}} > M_{\chi_1^+} > M_{\tilde{\ell}/\tilde{\nu}}>M_{\chi_1^0}$. 
Overall mass difference $M_{\tilde{t}} - M_{\chi_1^0} \equiv  \dmtotal$ smaller than the top quark
mass $M_t$ ensures that the uncompressed two body decay mode $\tilde{t}\;\to\;t\,\chi_1^0\;$ is 
kinematically forbidden.
Moreover, individual intermediate mass gaps {\it i.e.}  
$M_{\tilde{t}} - M_{\chi_1^+} \equiv \dmone$ and  $M_{\chi_1^+} - M_{\tilde{\ell}/ \tilde{\nu}} =
M_{\tilde{\ell}/ \tilde{\nu}} - M_{\chi_1^0} \equiv \dmtwo$ are also small due to the compressed nature of the considered simplified model. We would vary these mass differences in different range and discuss the consequence in subsequent sections.
It is evident that the following relation always holds: 
\be
\dmtotal = \dmone + 2 \times \dmtwo\label{deltTotal},
\ee
where we have assumed that the slepton and sneutrino are mass degenerate and lie exactly in the middle of mass gap between chargino and LSP.	
These assumptions are made to ensure an enhanced production of signal events and avoiding the possible very soft production of final state particles leading to efficient detection.
The branching fraction of wino-like chargino  $\chi_1^+$ decaying to slepton or sneutrino 
depends on the diagonalizing matrices.  %
However, since the second chargino is very heavy, we 
assume that the $\chi_1^+$ decays democratically to  both sleptons and sneutrinos i.e. 
BR($\chi_1^+\to\ell\tilde{\nu}$) = BR($\chi_1^+\to\nu\tilde{\ell}$) = 50\%.

One important feature of the assumed spectrum is that the top squark pair decays to dileptons without any substantial branching ratio suppression, whereas in the conventional compressed SUSY case the top squark pair decays to dileptons  via off-shell decay of the $W$ boson which has relatively small branching ratio to leptons($\sim 20\%$). Thus, the particular channel that we consider in this work is capable of delivering strong limit on top squark mass which in turn can constrain chargino and sleptons/sneutrinos mass even within the framework of compressed spectrum. Here we want to emphasize that though the top squark proceeds via sequential two body decay, the phase space available to each decay is highly compressed because of the two additional resonant states present in between the top squark and LSP.
In addition to the cross-section enhancement, the kinematic variables, which are suitably constructed for such scenario, are also very effective in this region because of the high level of compression in the spectrum. Now we describe these variables before utilizing them in our analysis in the next section.
%

We considered two new and simple kinematic variables which are very effective in the compressed region in discriminating signal and background. %
These variables were earlier proposed in our previous work~\cite{Konar:2016ata} where we have studied the four body decay of the top squark in the dileptonic channel. These variables are defined as the ratio between vector sum of all $b$ jet (lepton) transverse momentum and missing transverse momenta as follows,
\begin{eqnarray}
  R_{b E} = \frac{|\sum \vec{P}_T^{b_i}|}{\met}, \; \; \; \; \; 
  R_{\ell E} = \frac{|\sum \vec{P}_T^{{\ell}_i}|}{\met}.
\end{eqnarray}
These variables are proved to be useful to distinguish the compressed top squark from the large $t\bar{t}$ background.
Now allowing electroweak states within such a narrow mass gap for a successive two body decay, 
we argue that these ratio variables can still offer as good as, if not better, discrimination compare to the conventional scenario.

In our present decay chain as shown in figure~\ref{fig:decay-chain}, when the top squark decays to $b\chi_1^\pm$ if the mass difference is such that 
$M_{\tilde t} = M_b +M_{\chi_1^\pm}$ then both the b-quark and the chargino  have the same velocity vector at the equal-velocity limit. Heavier top squarks dominantly produced at the threshold and there by originate soft decay products if the mass difference are small.
However, if the top squark recoils against a high $p_T$ ISR then the decay particles will have moderate transverse energy and owing to the 
equal velocity approximation we can write 
\be \label{eq:equal-velocity}
p_T(b) = \frac{M_b}{M_{\tilde t}} p_T(\tilde t), \hspace{0.5cm} 
p_T(\chi_1^\pm) = \frac{M_{\chi_1^\pm}}{M_{\tilde t}} p_T(\tilde t).
\ee
Furthermore if we assume the approximation holds for the subsequent decay of chargino 
and slepton then the following relation holds:
\be\label{eq:more-equal-velocity}
p_T(\tilde \ell) = \frac{M_{\tilde \ell}}{M_{\chi_1^\pm}} p_T(\chi_1^\pm) 
\hspace{0.5cm} \textrm{ and }\hspace{0.5cm} 
p_T(\chi_1^0) = \frac{M_{\chi_1^0}}{M_{\tilde \ell}} p_T(\tilde \ell).
\ee
Using the equation~\ref{eq:equal-velocity} and \ref{eq:equal-velocity}  we have,
\bea
\sum p_T(b_i) &=& \frac{M_b}{M_{\tilde t}} \left( p_T(\tilde t) + p_T({\tilde t}^*)\right) 
              = \frac{M_b}{M_{\tilde t}}\cdot \frac{M_{\tilde t}}{M_{\chi_1^\pm}}  \sum p_T(\chi_1^\pm) \nonumber \\
	       &=& \frac{M_b}{{M_{\chi_1^\pm}}}
	                 \cdot \frac{M_{\chi_1^\pm}}{M_{\tilde \ell}} \sum p_T(\tilde \ell) 
		= \frac{M_b}{M_{\tilde \ell}}\cdot \frac{M_{\tilde \ell}}{M_{\chi_1^0}} \sum p_T(\chi_1^0),
\eea
where $\sum p_T(\chi_1^0)$ is the dominant component of the missing energy $\met$. The same exercise can be done 
for  $\sum p_T(\ell_i)$ which results into $\frac{M_\ell}{M_{\chi_1^0}} \sum p_T(\chi_1^0)$. Above arguments should suffice to demonstrate that the consecutive application of equal-velocity approximations ultimately cancel the mass dependence of intermediate states, so that the final algebraic relation remains as proposed in~\cite{Konar:2016ata}. Evidently these variables  are well suited to study the signature considered here.

In the present study the phase space available for  the $b$-jet and the lepton  
is determined by the mass gaps $\dmone$ and $\dmtwo$ respectively. Although the exact mass relation for equal velocity approximation does not 
hold, since the phase space is completely determined by the mass gaps,
the information can be used judiciously for the demarcation of signal and background events. 
For the background events, the  b-jets are originated from the top decay and the leptons stemmed  
from the decay of an on-shell $W$ boson. We demonstrate that although the variables 
$R_{b E}$ and $R_{\ell E}$ were initially proposed for the search of the four body decay region of the 
top squark, these variables are versatile enough to be used in different search channels like 
the present scenario at the LHC with the compressed spectrum.

In addition to the $R_{b E}$ and $R_{\ell E}$, the invariant mass variable between $b$ and $\ell$ system also encodes the mass differences between the  mediators. When the top squark decays via sleptons as shown in figure~\ref{fig:decay-chain}(A),
then the endpoint of the $b\ell$ system invariant mass is given 
by~\cite{Miller:2005zp}
\be\label{eq:mbl-slepton}
\left(M_{b\ell}^{max}\right)_{\tilde{\ell}}^2 = \frac{\left(M_{\tilde{t}}^2-M_{\chi_1^+}^2\right)\left
                                        (M_{\tilde{\ell}}^2-M_{\chi_1^0}^2\right)}{M_{\chi_1^+}^2}.
\ee
On the other hand if there is sneutrino in the mediator, as in figure~\ref{fig:decay-chain}(B), then the invariant mass 
distribution of $b\ell$ system have an endpoint at~\cite{Miller:2005zp}
\be\label{eq:mbl-sneutrino}
\left(M_{b\ell}^{max}\right)_{\tilde{\nu}}^2 = \frac{\left(M_{\tilde{t}}^2-M_{\chi_1^+}^2\right)\left
  (M_{\chi_1^+}^2-M_{\tilde{\nu}}^2\right)}{M_{\chi_1^+}^2}.
\ee
Hence, depending on the mediator the endpoint will change for a particular $b\ell$ combination. 
For a given spectrum with degenerate slepton  and  sneutrino masses, it is evident from above equations that 
the sneutrino mediated quantity is always larger than the other, 
{\em i.e.} $\left(M_{b\ell}^{max}\right)_{\tilde{\nu}} > \left(M_{b\ell}^{max}\right)_{\tilde{\ell}}$. In a compressed scenario like we have considered here, this difference is typically below a GeV, and we estimate the theoretical endpoint $M_{b\ell}^{max}$ as a event selection parameter by considering the largest of these two. It is worth noting that the variable $R_{\ell E}$ and $M_{b\ell}$ are quite sensitive to the mass gap between the intermediate resonances. Moreover, $R_{\ell E}$ is sensitive only to $\dmtwo$ while $M_{b\ell}$ depends on both $\dmone$ and $\dmtwo$.
 
The background for the dilepton channel with at least one b-tagged jets comes predominantly from 
$t\bar{t}$ with sub-leading contribution comes from $tW$ channel. The $t\bar t$ background decay 
chain is shown in right panel of figure~\ref{fig:decay-chain}. For the background process the 
$b\,\ell$ invariant mass has a endpoint at $\sqrt{M_t^2 - M_W^2} \simeq 153$ GeV which is way 
larger than the $M_{b\ell}^{max}$ for signal events in the compressed spectrum. 
In the next section we show how these variables can be used to minimize the backgrounds.

\section{Analysis setup and Simulation}\label{simulation} %
We consider the pair production of top squark associated with a hard (ISR) jet against which the top squark pair recoils. We are looking into the following  signal process,
\be\label{eq:sig}
P P \rightarrow 2 \ell \;+\; \ge 1 \text{ b-tagged-jet} \;+\;  \ge 1 \text{ high-PT jet (ISR)} \;+\;  \met.
\ee
To simulate the signal events we assume a simplified model with the mass hierarchy 
$M_{\tilde{t}} > M_{\chi_1^+} > M_{\tilde{\ell}/\tilde{\nu}}>M_{\chi_1^0}$ and two body successive decay to produce above mentioned signature.
The signal events as well as the background events are generated  using  
$\texttt{MadGraph5\_aMC@NLO}$~\cite{Alwall:2014hca} with \texttt{NN23LO1} \cite{Ball:2014uwa} parton 
distribution function and with the default dynamic 
renormalisation and factorisation scales \cite{mad:scale} in $\texttt{MadGraph5\_aMC@NLO}$. 
\texttt{Pythia8}~\cite{Sjostrand:2006za,Sjostrand:2007gs}is used for showering and hadronization 
of the parton level events and subsequently full detector level simulation is done in 
\texttt{Delphes3}~\cite{deFavereau:2013fsa}. All the samples are matched up to 
one jets using MLM scheme\cite{Mangano:2006rw,Hoche:2006ph}. For all our analysis we have used 
the next-to-leading-order (NLO) plus next-to-leading-logarithm (NLL) top squark cross-sections 
given by the LHC SUSY cross-section Working Group~\cite{LHC_stop_xsection,Borschensky:2014cia}.

The SM processes which can mimic the signal can come from $t\bar t$ and $tW$ processes where the top decays leptonically. We have also simulated these background events using  
$\texttt{MadGraph5\_aMC@NLO}$ followed by \texttt{Pythia8} and \texttt{Delphes3} and are matched 
up to one jets using MLM scheme. The $t\bar{t}$ production cross-section 
is $\sigma_{t\bar{t}}$ = 815.96 pb as calculated with the Top++2.0 program to 
next-to-next-to-leading order in perturbative QCD, including soft-gluon resummation to 
next-to-next-to-leading-log order (see \cite{Czakon:2011xx} and 
references therein), and assuming a top-quark mass $m_{t}$ = 173.2 GeV. 
For the $tW$ channel the NLO+NNLL cross-section is 71.7 pb~\cite{Kidonakis:2015nna}. It is 
obvious that the dominant contribution to the background events are coming from the 
$t\bar t$ events due to large production cross-section.

The jets are reconstructed using the  anti-kT algorithm \cite{Cacciari:2008gp} 
with jet radius $R = 0.4$. We tagged the highest $p_T$ jet as the ISR jet and we require $p_T(ISR) > 100$ GeV. 
A moderately high ISR ensures that the pair produced top squarks recoil against the ISR and  
produce large missing transverse energy. In order to minimize the QCD background we use the fact that 
the ISR and MET are in opposite direction for signal events and we apply the cut $|\Delta\phi(\met,ISR)| < 0.5$
where $\Delta\phi(\met,ISR) = \phi(\met)-\phi(ISR)-\pi$.  
We also demand $|\phi(j)-\phi(\met)| > 0.2$ for all jets other than the ISR in order 
to minimize the effect of jet mis-measurement contributing into $\met$.

For b-jet tagging  we use the Combined Secondary Vertex algorithm 
at the Medium operating point(CSVM) since the tagging 
algorithm delivers the best efficiency at medium operating point~\cite{Chatrchyan:2012jua}. 
The b-tagging efficiency depends on the transverse momentum of the 
jet and parametrized in \texttt{Delphes3} as, 
\be
\mathcal{E}_{b-tag} = 0.85\tanh\big[0.0025\;p_T(j)\big] \frac{25.0}{1+0.063\times p_T(j)},
\ee
where $p_T(j)$ is the jet transverse momentum and $p_T(j) \geq 20$ GeV. The tagging efficiency 
is close to 50\% for low $p_T$ jets and peaks at $p_T(j) \simeq 150$ GeV where 
$\mathcal{E}_b \simeq  73\%$. Mis-tagging efficiency of the light jets is parametrized as 
\be
\mathcal{E}_{mis-tag} =  0.01 + 0.000038 \;p_T(j)
\ee
which increases linearly and becomes 1.5\% for $p_T(j) \simeq 150$ GeV. Since we are interested in the compressed spectrum, the b-jets tend to have low $p_T$ and hence due to less efficiency it is not economical to tag both the b-jets.   

\begin{figure}[t]
 \centering
 \includegraphics[keepaspectratio=true,scale=0.6]{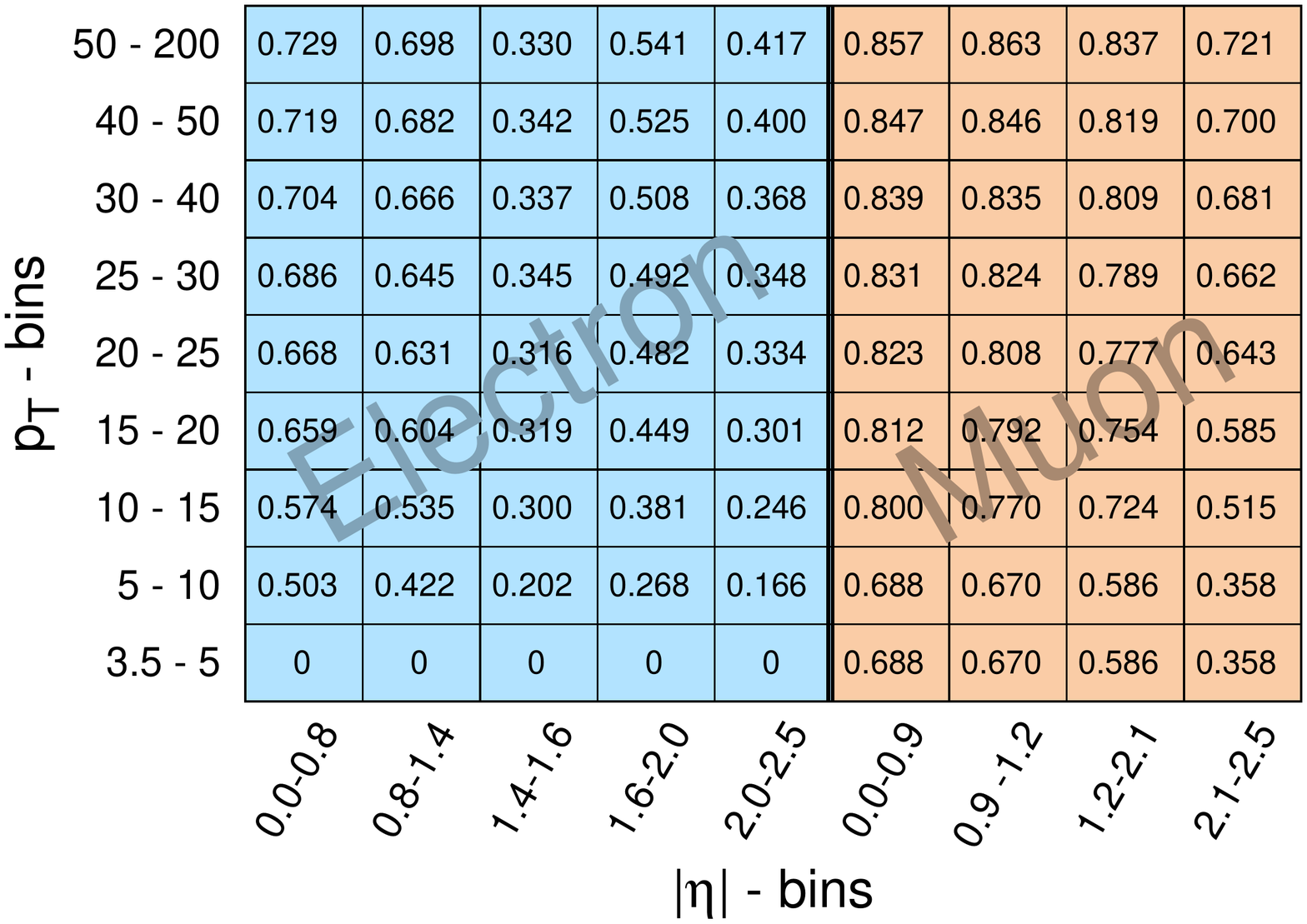}
 \caption{Tagging efficiencies for leptons in different $p_T-\eta$ bin\cite{light-lepton-efficieny}. 
   Minimum $p_T$ for electron is 5 GeV and for muons it is 3.5 GeV. The efficiencies 
   for very soft leptons are small and overall efficiency to tag an electron 
   is less than that of muons. }
 \label{fig:lep-tagging}
\end{figure}

The transverse momentum of the leptons depends on the mass differences $\dmtwo$ and can be 
very soft. The CMS collaboration has studied the prospect of soft leptons in compressed top squark 
search~\cite{CMS-PAS-SUS-16-025} where varying light lepton selection efficiency was used 
depending on the transverse momentum and pseudorapidity of the leptons~\cite{light-lepton-efficieny}.
The electron should have a minimum $p_T(e)$ of 5 GeV, whereas the muons are required to have $p_T(\mu)$ 
at least 3.5 GeV. For a realistic simulation we have used the exact lepton identification efficiencies 
used by the CMS collaboration~\cite{light-lepton-efficieny}. The exact values of efficiencies are 
shown in figure~\ref{fig:lep-tagging} for different $p_T - \eta$ bin. As one can see the efficiencies 
for very soft leptons are small and overall efficiency to tag an electron is smaller than that of muons. 
\begin{figure}
 \centering
 \includegraphics[keepaspectratio=true,scale=0.8]{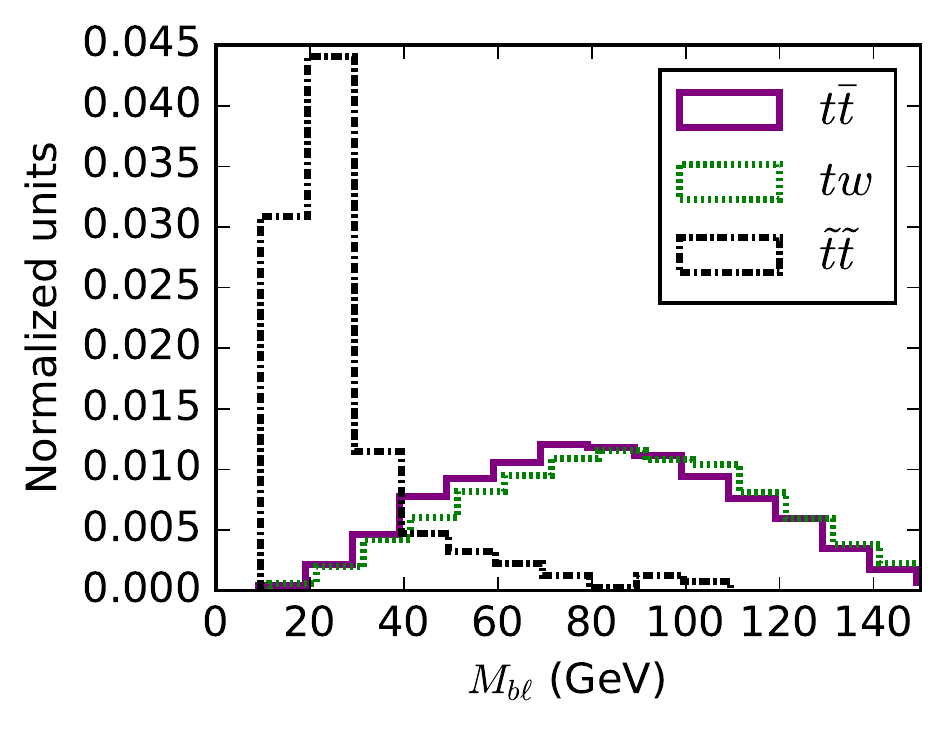}
 \includegraphics[keepaspectratio=true,scale=0.8]{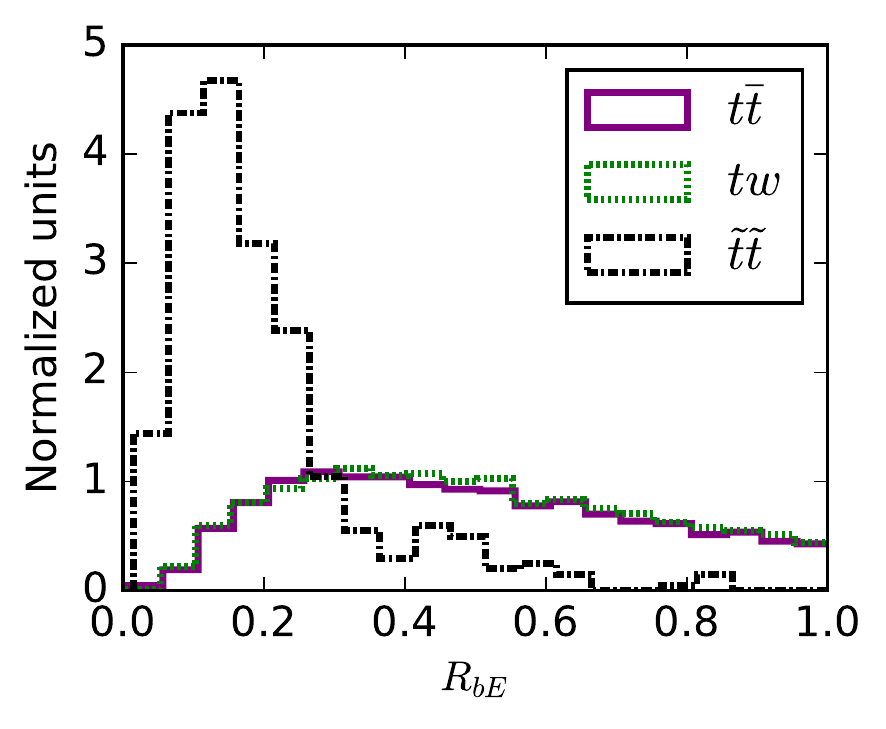}
 \includegraphics[keepaspectratio=true,scale=0.8]{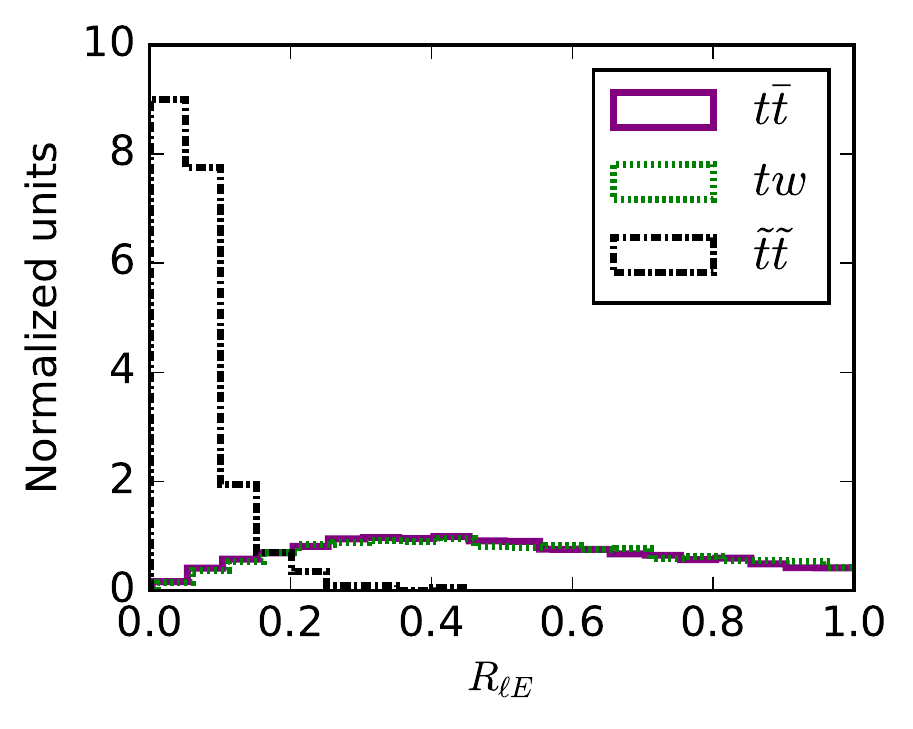}
 \includegraphics[keepaspectratio=true,scale=0.8]{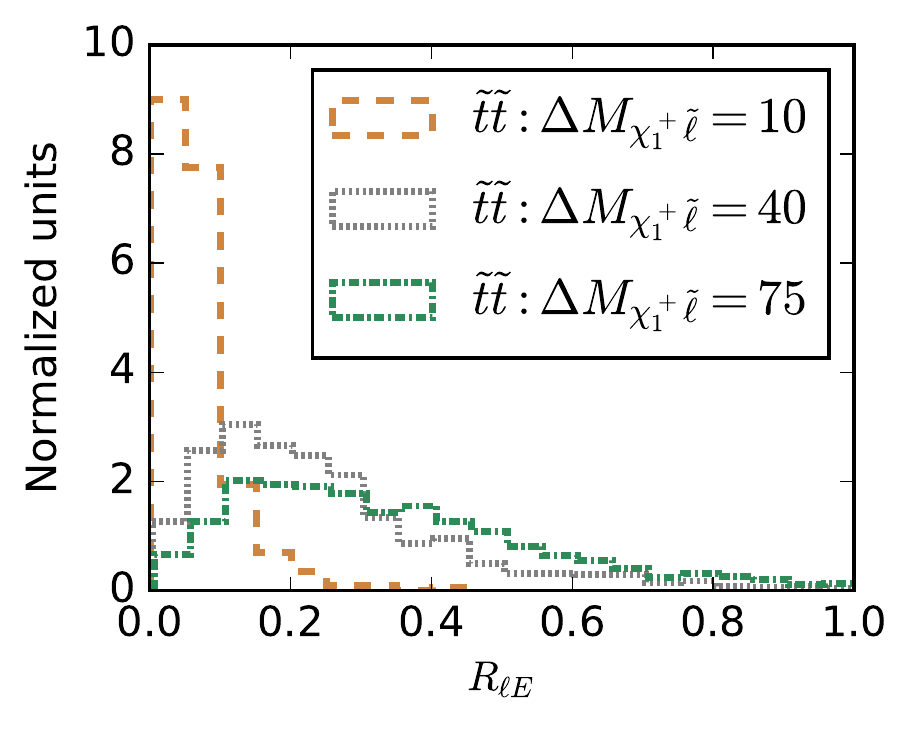}
 \caption{Normalized distributions of different relevant kinematic variables $M_{b\ell}$, $R_{b E}$ and $R_{\ell E}$ which are
 used in the present analysis to get improved signal over the background. Black (dash-dot) distribution lines in each plot is for signal top squark pair production, along with two dominant background processes. 
 All the plots except the bottom right one are generated for a representative benchmark point with $\dmone = 20$ GeV and 
 $\dmtwo = 10$ GeV. The bottom right plot illustrates the variation on the signal distribution on the  mass difference $\dmtwo$ in GeV.
 For all the distributions it is easy to observe that the background and signal processes are well separated and the 
 lower the $\Delta M_{ij}$, the bigger the separation between signal and backgrounds.}
 \label{fig:variable-distribution}
\end{figure}

The kinematics of the leptons and b-jet depend strongly on mass gaps $\dmtwo$ and $\dmone$.  
Since the minimum transverse momentum required by a tagged b-jet needs to have 20 GeV, the minimum $\dmone$ should be close or greater than the same to allow enough phase space for the $b-$ quark. In similar argument we set the benchmark points such that the $\dmtwo$, 
 which controls the lepton $p_T$, required to have minimum value of 10 GeV.    
 Using the present search strategy it is extremely difficult to search compressed 
spectrum below the minimum values of $\Delta M_{ij}$ chosen here and we may need different strategy like 
search for long lived charged tracks etc.  From these minimum values of $\Delta M_{ij}$ it is evident that 
we can not search for compressed region where overall mass difference $\dmtotal$ is less than 40 GeV (see equation.~\ref{deltTotal}). Moreover, b-jet tagging efficiency is poor for small $P_T$ regions which we are actually interested to probe and hence our method can only probe stop-chargino mass gap above 20 GeV (as evident in Fig.~\ref{fig:chargino_limit-plot}). Any further improvements in low $P_T$ b-tag can enable us to probe low mass gap region. 
Secondly, $\met$ is calculated indirectly from the transverse momentum imbalance of all the isolated objects such as leptons, photons and jets, as well as any unclustered deposits. Accuracy of this measurement is susceptible to underlying events especially the pile up in high luminosity.

The phase space available for the leptons depend on the mass gap $\dmtwo$ and consequently the 
$R_{\ell E}$ distribution varies with it.  We found that as $\dmtwo$ increases the 
$R_{\ell E}$ distribution move towards higher values and upper edge of the distribution follows 
the following empirical relation:
\be\label{eq:rle-variation}
R_{\ell E}^{max} = 0.0576 +0.0047 \; \frac{\dmtwo}{GeV}.
\ee
We have used this relation to find $R_{\ell E}^{max}$ for a particular $\dmtwo$ in each benchmark point. 
This criteria is very effective  in
minimizing the backgrounds as for the SM processes the leptons are originated from the on-shell $W$ 
boson and tends to have higher values of $R_{\ell E}$. Similarly, $R_{b E}$ distribution depends on $\dmone$ but the dependency is mild. To keep our presentation simple for the present analysis we fix it to a constant value $R_{b E}^{max} = 0.2$. 

In figure~\ref{fig:variable-distribution} we have presented different variables which are used to 
minimize the background. For this demonstration, all the plots are generated considering $\dmone = 20$ GeV and 
 $\dmtwo = 10$ GeV. While these distributions are are independent of the top squark mass, here we took its value as 700 GeV.
 The bottom right plot illustrates the variation on the distribution of 
 $R_{\ell E}$ based on the mass difference $\dmtwo$. As $\dmtwo$ increases, the transverse 
 momentum of the leptons also increases leading to a larger value of $R_{\ell E}$. From the distributions
 it is quite comprehensible that the background and signal processes are well separated and the 
 lower the $\Delta M_{ij}$, the bigger the separation between signal and backgrounds. This leads to better 
 signal to background ratio for lower mass gap regions. However the lower mass gap region is inherently 
 adverse due to lower signal acceptance factor (since the final state particles are soft) which to some 
 extend normalizes the improvement achieved from the well separated signal and backgrounds.
 Top left panel of figure~\ref{fig:variable-distribution} represents the invariant mass of $b$-jet and lepton.  Since we have tagged only one $b$-jet, there are two possible $b\ell$ combination and we choose the smaller value among those two to calculate the $M_{b\ell}$ . For the given $\dmone$ , $\dmtwo$ and the top squark mass in the shown example, $M_{bl}^{max}$ turns out to be around 28 GeV where difference between two equations is in sub-GeV.

\section{Result and Discussions}\label{results}

  \begin{table}[t]
 \centering
 \begin{tabular}{|c||c||c|c||c|}
 \hline
  Cut                             & {Signal}        & \multicolumn{2}{c||}{Background}     & {Significance}   \\
                                  &                 & $t\,\bar{t}$      & $t\,W$          & $@300\rm{fb}^{-1}$ \\ \hline
 Pre-selection+&\multirow{2}{*}{371 [$100\%$]}&\multirow{2}{*}{348680 [$100\%$]}&\multirow{2}{*}{40090 [$100\%$]}&\multirow{2}{*}{0.6}\\ 
 $2\ell + ISR+\geq 1\,b$          &                 &                   &                 &   \\ \hline
 $\met >$ 200 GeV                 & 177  [$47.8\%$] & 17280  [$4.9\%$]   & 3464  [$8.6\%$] &1.2\\ \hline
 $|\Delta\phi(\met,ISR)| < 0.5$ & 153  [$41.3\%$] & 12840  [$3.7\%$]    & 2364  [$5.9\%$] &1.2  \\ \hline
 $M_{b\ell} < 50$                 & 130  [$35.2\%$] & 2661  [$0.8\%$]    & 277  [$0.7\%$]  &2.4 \\ \hline
 $R_{bE} < 0.2$                   & 112  [$30.2\%$] & 1415  [$0.4\%$]    & 155  [$0.4\%$]   &2.8 \\ \hline
 $R_{\ell E} < 0.152$             & 100  [$27.0\%$] & 325  [$0.09\%$]     & 19  [$0.05\%$]  &5.2 \\ \hline
 \end{tabular}
 \caption{The effectiveness of different variables to minimize the colossal background is shown in this 
   cutflow table. This representative signal is generated for $\Delta M_{\tilde{t}\chi_1^+} = 30$ GeV and 
$\Delta M_{\chi_1^+\tilde{\ell}}=20$ GeV with the fixed the top squark mass at 725 GeV. 
All the number of events for signal and background are generated with integrated luminosity of 300 $\rm{fb}^{-1}$. 
The maximum qualifying values for $M_{b\ell}$ and $R_{\ell E}$ are selected according to the mass gaps as described in the text. $R_{bE}^{max}$ is kept constant for all benchmark points. As one can see due to narrow mass gap in the mass 
spectra the variables efficiently retain at least 27$\%$ of the signal events while discarding more than 
99.9$\%$ of the background events leading to 5$\sigma$ significance.}
 \label{tab:cut-flow}
 \end{table}
In previous sections we discussed the model framework and characteristic kinematic distributions which can be utilized to improve the signal to background ratio.
To quantify the efficacy of different proposed kinematic variables we consider a representative benchmark point with $\dmone = 30$ GeV and $\dmtwo = 20$ GeV, {\it i.e.} total mass gap between top squark and the lightest neutralino is below the $W$ boson mass $M_W$.  Detailed step by step cut flow is presented in table~\ref{tab:cut-flow} corresponding to this signal point along with the backgrounds. The events are estimated with an integrated luminosity of $300~\rm{fb}^{-1}$ and the signal events are generated with top squark mass 725 GeV.
As mentioned before, the exact choice of  $M_{b\ell}$ selection criteria is decided by the mass gaps as in equations~\ref{eq:mbl-slepton} and~\ref{eq:mbl-sneutrino}. 
Similarly, the maximum allowed value of $R_{\ell E}$ varies according to 
equation~\ref{eq:rle-variation} primarily depending on $\dmtwo$. The choice of  
$R_{bE}$ cut also, in principle, depends on $\dmone$. For the sake of simplicity, we fix the maximum allowed value of $R_{bE}$ at $0.2$ to select our events. With these sequential cuts we have analyze the signal ($S$) and background events ($B$),
and also present the corresponding statistical significance ($\mathcal{S}$) at each step in the right most column. We estimate the $\mathcal{S}$ 
using the well known expression:
\be
\mathcal{S} = \sqrt{2\left[(S+B)~\textrm{ln}\left(1+\frac{S}{B}\right)-S\right]}.    %
\ee

Since the presence of lightest neutralino in the 
final state results larger $\met$ for signal events compared to the background, we have chosen minimum MET 
of 200 GeV which minimizes the background considerably. However, the dominant background cross-section is 
three-orders of magnitude higher than the signal cross-section and it is evident from the significance column that 
only MET cut (which is the usual search strategy for compressed spectrum) is not enough to size the background. 
As we introduce the new kinematic variables the background events drop rapidly, consequently the 
significance increases steadily and reaches 5$\sigma$. This cut flow table demonstrates the usefulness of 
the  proposed variables.

\begin{figure}
 \centering
 \includegraphics[keepaspectratio=true,scale=1]{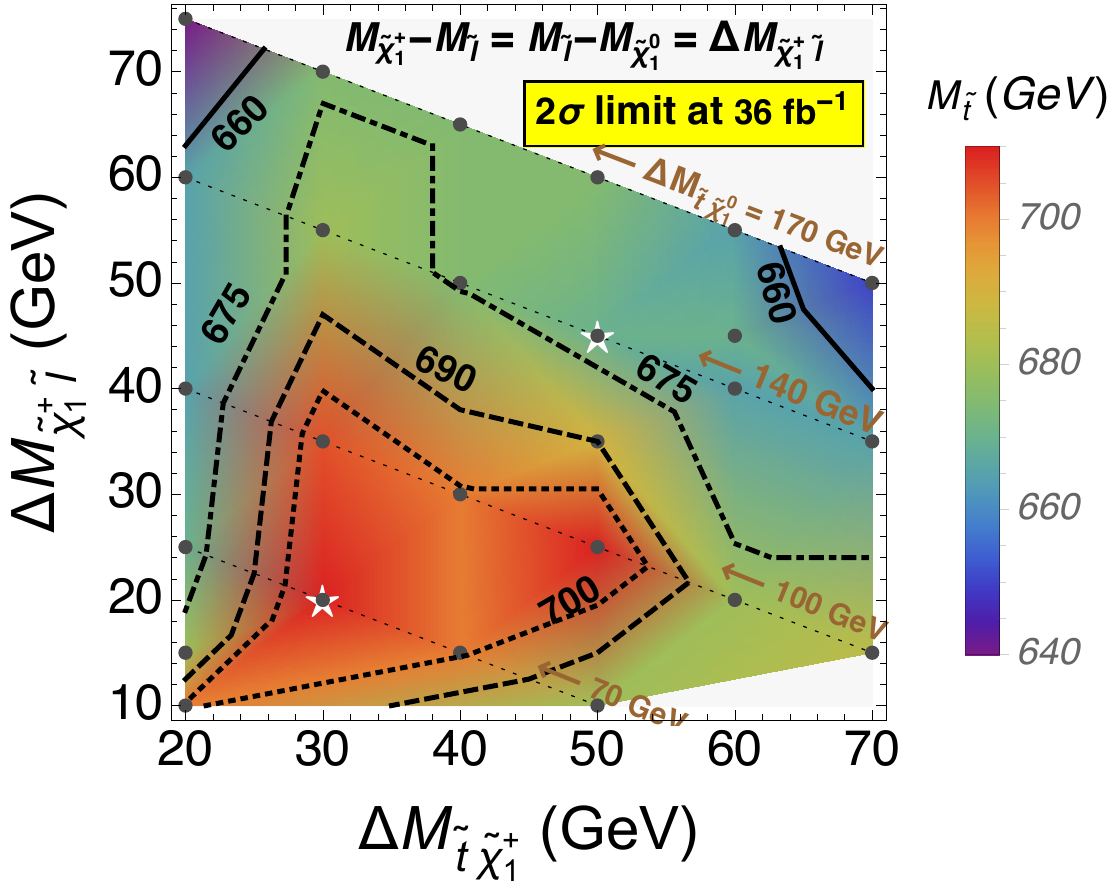}
 \caption{Reach of 13 TeV LHC with 36 $\rm{fb}^{-1}$ data for different values of different mass gaps between 
 top squark, chargino and the neutralino. The horizontal axis is responsible for the b-jet $p_T$, while the vertical 
 axis controls the lepton phase space. The white space in top right corner is not analyzed as in this region the 
top squark - neutralino mass difference goes beyond the top-corridor. The bottom right white space is numerical artifact 
due to computational constraints. }
 \label{fig:heatmap}
\end{figure}
Now to explore the full model parameter space in terms of two relevant mass differences $\dmone$ and $\dmtwo$, we vary both these mass gaps in such a way that the total mass difference between top squark and the lightest neutrino $\dmtotal$ does not go beyond the top quark mass. To achieve this we vary the mass difference $\dmone$ in between 20 GeV to 70 
GeV with a step of 10 GeV and $\dmtwo$ is chosen accordingly. 
The maximum allowed value of the parameters $M_{b\ell}$ and $R_{\ell E}$ changes according 
the mass gaps. 
Using the new kinematic variables we 
calculated the 2$\sigma$ exclusion limit at 13 TeV LHC with integrated luminosity of 36 $\rm{fb}^{-1}$ and is depicted in 
the heat plot in figure~\ref{fig:heatmap}. 
The abscissa is showing the mass gap between top squark and chargino $\dmone$
which is responsible for the b-jet phase space and the ordinate is half of the chargino neutralino 
mass difference $\dmtwo$ which controls the lepton $p_T$ spectrum. 
The color bar is showing the top squark mass which can be excluded for given mass gaps in both the axes. 
For readability, we put few contours with fixed top squark mass. The diagonal dotted lines represent a fixed top squark-neutralino mass gap which are indicated with $\dmtotal = 170,~140,~100$ and $70$ GeV. 
Along these dotted lines all the off-white points  are the benchmark points we have chosen to scan the parameter space. 
Among these benchmark points, two white stars marks are indicated for which we supplied the complete mass spectrum at the Appendix Table~\ref{tab:bp_comp}, satisfying all the collider, DM and low energy constraints.
All the benchmark points and the corresponding cut values of different variables are tabulated later in table~\ref{tab:benchmark}. 

It is evident from the figure~\ref{fig:heatmap} that the LHC reach is more sensitive to the vertical axis which 
regulates the leptonic phase space. As we move towards the higher values in vertical axis, the leptons 
tend to have larger transverse momentum which increase $R_{\ell E}^{max}$. Since the dominant $t\bar{t}$ 
background tends to have $R_{\ell E}$ distribution which peaks at a higher value, increase in 
$R_{\ell E}^{max}$ results in decreasing the LHC reach. For different values of the horizontal axis  the 
b-tagged jet $p_T$ varies its effect in the final significance is mild. However, it is obvious that our 
proposed variables works well at the lower end of the mass difference which is very difficult region to explore. The white region in the top right corner has not been explored in this work since we 
adhere to the fact that the top squark neutralino mass difference should not go beyond the so called top-threshold, 
beyond which the 2-body top squark decay $\tilde{t}\to t\,\chi_1^0$ is important and should be taken into 
account to achieve a realistic limit on the top squark mass.

\begin{figure}
 \centering
 \includegraphics[keepaspectratio=true,scale=0.45]{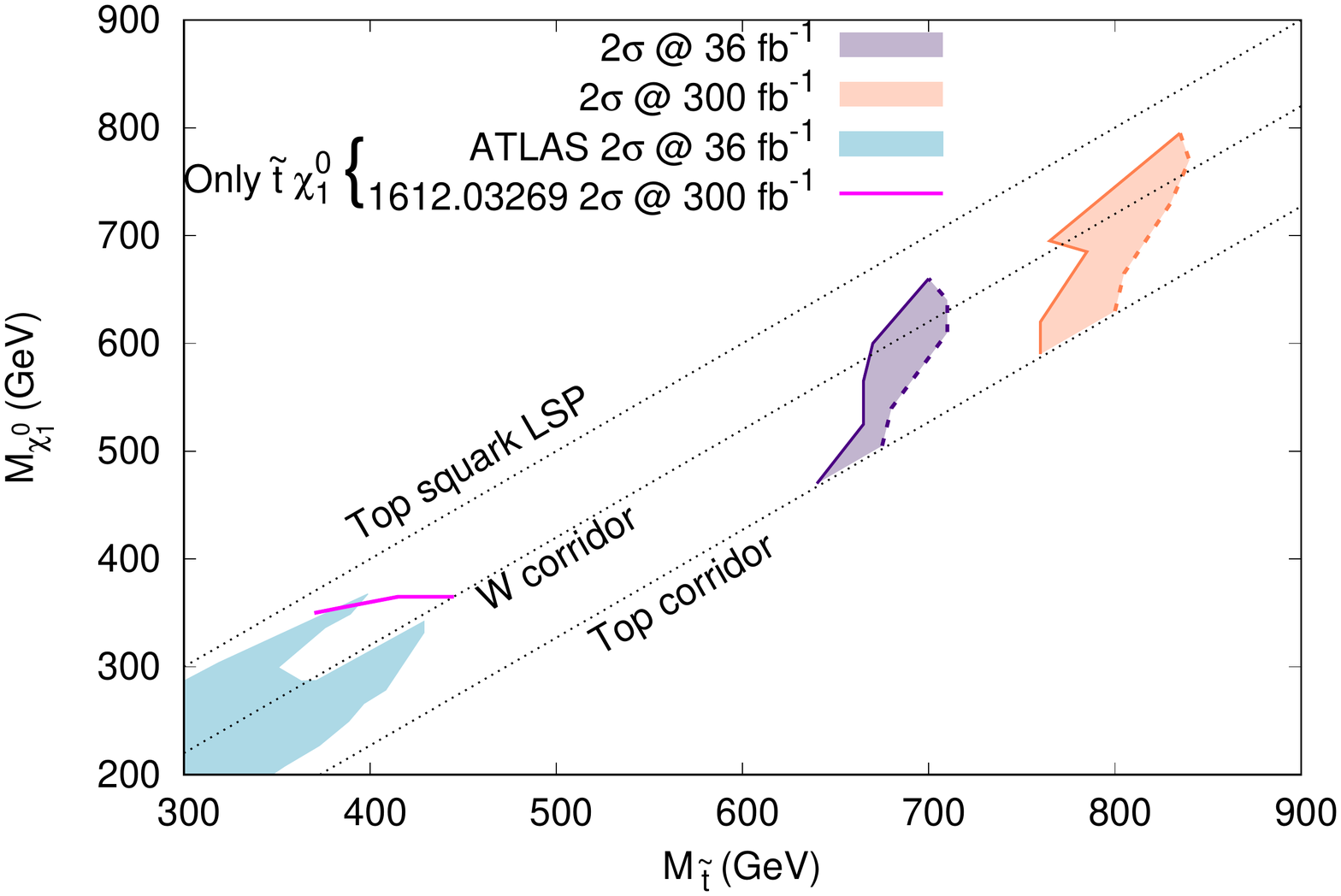}
 \caption{$95 \%$ exclusion limit in $M_{\tilde{t}}$ - $M_{\chi_1^0}$ plane considering integrated luminosity as 
     36 $fb^{-1}$  and 300 $fb^{-1}$ with light purple shade and light orange shade respectively. Limits are 
     obtained from  tri-lepton + $\met$ final state originated from $\tilde{t}$ decay with $\chi_1^\pm$, $\chi_2^0$ as 
     intermediate states.
     }
 \label{fig:limit-plot}
\end{figure}

To further understand the implication of the present bound, we have plotted in figure~\ref{fig:limit-plot} 
the exclusion limits in the  $M_{\tilde{t}} - M_{\chi_1}^0$  plane along with other %
exclusion limits obtained from different compressed searches. The purple bold line depicts the minimum top squark 
mass which can be excluded at 2$\sigma$ at 13 TeV LHC with integrated luminosity of 36 $\rm{fb}^{-1}$ 
data for a given top squark-neutralino mass gap provided $\dmone$ and $\dmtwo$ lie within their range of our described benchmark points\footnote{We already clarified earlier that the extreme small choice of any of these $\Delta M_{ij}$ can push this lower line anywhere below by making the analyzed channel undetectable in present search. Hence this band represents between most optimistic scenario to the extend this signature is promising at the LHC~\cite{Konar:2010bi}.}.
The best exclusion limit on top squark mass is achieved for a particular 
$\dmone$ and $\dmtwo$ combination which is the most sensitive to the kinematic variables.
This limit is shown by the purple dotted line. 
For clarification one can follow from the heat map plot in figure~\ref{fig:heatmap}.
The best limit on top squark mass is 710 GeV and is achieved for $\dmone = 30$ GeV and $\dmtwo = 20$ GeV, whereas
the minimum limit is 640 GeV and with $\dmone = 20$ GeV and $\dmtwo = 75$ GeV. They represent two points over these 
dashed and solid purple curves respectively.

Anything between these two lines can be excluded for various $\dmone$ and $\dmtwo$ combination and are shown with light purple hue.
Similarly orange curves and the light orange color band is presented for  2 $\sigma$ exclusion limit considering 300 $fb^{-1}$  integrated luminosity.
In the same figure,
current limit on top squark search through dilepton channel is shown in light blue patch as reported 
by the ATLAS collaboration~\cite{ATLAS-CONF-2017-034}. In our previous work~\cite{Konar:2016ata} we have studied the utility of the kinematic 
variables $R_{\ell E}$ and $R_{b E}$ for the compressed scenario where the top squark undergoes a four-body 
final state decay. The limit on top squark mass with data equivalent to integrated luminosity of 300 $\rm{fb}^{-1}$ 
is shown in magenta curve . 
However these two searche limits assume that top squark to be the NLSP in contrary to 
the spectrum we are concerned here : $M_{\tilde{t}} > M_{\chi_1^+} > M_{\tilde{\ell}/\tilde{\nu}}>M_{\chi_1^0}$ 
and these limits are shown for reference only. Effective kinematic variables and absence of leptonic $W$ boson 
decay in the signal process help to achieve a higher limit in the present search channel. 
%

\begin{table}[t!]
	\centering
	\begin{tabular}{|c|c||c|c|c||c|c|c|}
		\hline
		\multicolumn{2}{|c||} {Parameters} &\multicolumn{3}{c||} {Kinematic Variables}&\multicolumn{3}{c|} {Limit on top squark mass (GeV)} \\
		\multirow{2}{*}
		{\begin{tabular}
				[c]{@{}c@{}}
				$\Delta M_{\tilde{t}\chi_1^+}$
				\\ (GeV)
			\end{tabular}} & \multirow{2}{*}
			{\begin{tabular}[c]{@{}c@{}}
					$\Delta M_{\chi_1^+\tilde{\ell}}$ 
					\\ (GeV)\end{tabular}}    & \multirow{2}{*}
			{\begin{tabular}[c]{@{}c@{}}
					$M_{b\ell}^{max}$\\
					(GeV)\end{tabular}} & \multirow{2}{*}
			{$R_{\ell E}^{max}$} & \multirow{2}{*}
			{$R_{b E}^{max}$}     & \multicolumn{3}{c|}
			{}  \\  
			&                        &                          &                  &               & $2\sigma\; @  36 \;
			\rm{fb}^{-1}$ & $2\sigma \;@ 300\;
			\rm{fb}^{-1}$ & $5\sigma \; @300\;
			\rm{fb}^{-1}$ \\ \hline\hline
			\multirow{6}{*}{20} & 10                    & 28                       & 0.105            & \multirow{6}{*}{0.2}  & 700   & 835         & 715  \\ \cline{2-4} \cline{6-8} 
			& 15                    & 35                       & 0.126            &                       & 680   & 815         & 700  \\ \cline{2-4} \cline{6-8} 
			& 25                    & 45                       & 0.175            &                       & 670   & 765         & 685  \\ \cline{2-4} \cline{6-8} 
			& 40                    & 58                       & 0.246            &                       & 665   & 785         & 680  \\ \cline{2-4} \cline{6-8} 
			& 60                    & 72                       & 0.339            &                       & 665   & 790         & 680  \\ \cline{2-4} \cline{6-8} 
			& 75                    & 81                       & 0.410            &                       & 640   & 760         & 655  \\ \cline{1-4} \cline{6-8} 
			\hline\hline
			\multirow{4}{*}{30} & 20                    & 50                       & 0.152            & \multirow{4}{*}{0.2}  & 710   & 840         & 725  \\ \cline{2-4} \cline{6-8} 
			& 35                    & 67                       & 0.222            &                       & 705   & 830         & 720  \\ \cline{2-4} \cline{6-8} 
			& 55                    & 84                       & 0.316            &                       & 680   & 805         & 695  \\ \cline{2-4} \cline{6-8} 
			& 70                    & 96                       & 0.387            &                       & 675   & 795         & 690  \\ \cline{1-4} \cline{6-8} 
			\hline\hline 
			\multirow{4}{*}{40} & 15                    & 50                       & 0.125            & \multirow{4}{*}{0.2}  & 700   & 825         & 715  \\ \cline{2-4} \cline{6-8} 
			& 30                    & 71                       & 0.199            &                       & 700   & 830         & 720  \\ \cline{2-4} \cline{6-8} 
			& 50                    & 93                       & 0.293            &                       & 675   & 800         & 695  \\ \cline{2-4} \cline{6-8} 
			& 65                    & 107                      & 0.363            &                       & 675   & 795         & 685  \\ \cline{1-4} \cline{6-8} 
			\hline\hline
			\multirow{5}{*}{50} & 10                    & 46                       & 0.105            & \multirow{5}{*}{0.2}  & 680   & 805         & 695  \\ \cline{2-4} \cline{6-8} 
			& 25                    & 73                       & 0.175            &                       & 710   & 840         & 725  \\ \cline{2-4} \cline{6-8} 
			& 35                    & 85                       & 0.222            &                       & 690   & 820         & 705  \\ \cline{2-4} \cline{6-8} 
			& 45                    & 99                       & 0.269            &                       & 670   & 760         & 685  \\ \cline{2-4} \cline{6-8} 
			& 60                    & 116                      & 0.339            &                       & 675   & 800         & 695  \\ \cline{1-4} \cline{6-8} 
			\hline\hline
			\multirow{4}{*}{60} & 20                    & 72                       & 0.152            & \multirow{4}{*}{0.2}  & 680   & 805         & 695  \\ \cline{2-4} \cline{6-8} 
			& 40                    & 103                      & 0.246            &                       & 665   & 790         & 680  \\ \cline{2-4} \cline{6-8} 
			& 45                    & 105                      & 0.269            &                       & 670   & 790         & 685  \\ \cline{2-4} \cline{6-8} 
			& 55                    & 122                      & 0.316            &                       & 665   & 785         & 680  \\ \cline{1-4} \cline{6-8} 
			\hline\hline 
			\multirow{3}{*}{70} & 15                    & 67                       & 0.128            & \multirow{3}{*}{0.2}  & 685   & 810         & 700  \\ \cline{2-4} \cline{6-8} 
			& 35                    & 105                      & 0.222            &                       & 665   & 790         & 680  \\ \cline{2-4} \cline{6-8} 
			& 50                    & 126                      & 0.293            &                       & 650   & 765         & 665  \\ \hline
		\end{tabular}
		\caption{Different benchmark points analyzed in the present study characterized by two mass gaps $\dmone$ and $\dmtwo$ satisfying a compresses mass gap between LCP and LPS as, $\dmtotal = \dmone + 2 \times \dmtwo$. According to the mass differences various cuts are fixed (see, equations ~\ref{eq:mbl-slepton},~\ref{eq:mbl-sneutrino} and~\ref{eq:rle-variation} with following text) and are tabulated in the kinematic variables column. Using these cuts we have calculated the top squark mass
			reach at 13 TeV LHC with integrated luminosity of 36 $\rm{fb}^{-1}$ and 300 $\rm{fb}^{-1}$.}
		\label{tab:benchmark}
		
	\end{table}

\begin{figure}
\centering
\includegraphics[keepaspectratio=true,angle=-90,scale=0.45]{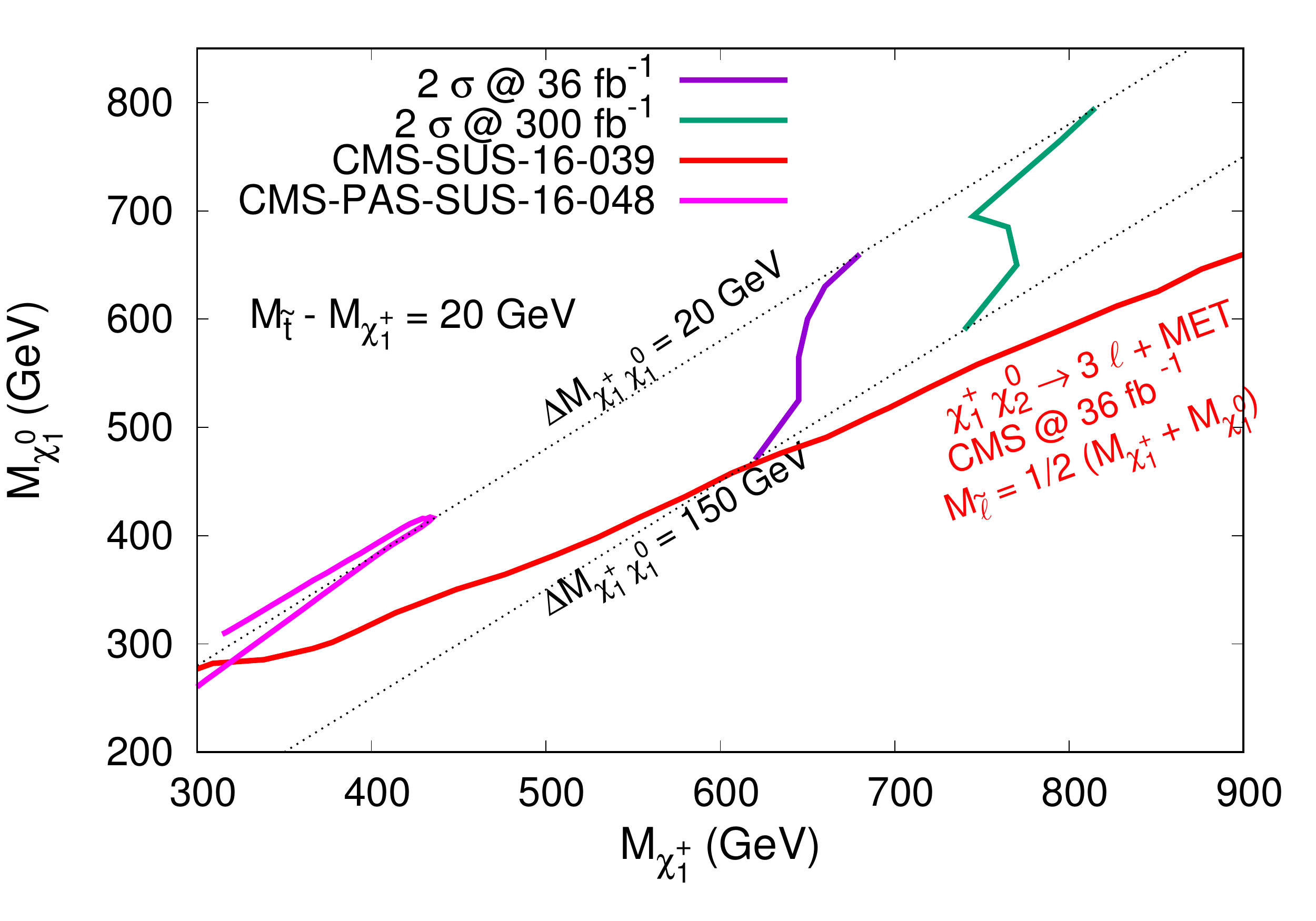}
\caption{Exclusion limit in chargino-neutralino plane at 13 TeV LHC considering compressed spectrum with top squark as lightest color sparticle. In order to show the efficacy of our  new observables we have presented the exclusion limits for chargino with $95 \%$ CLs corresponding to 36 $fb^{-1}$ luminosity as displayed by the purple line. The limits corresponding to 300 $fb^{-1}$ luminosity is shown by the green line. The red line depicts the current CMS search limit where they have studied the trilepton along with missing energy considering direct production of chargino-neutralino. 
  There is no experimental search in the dileptonic decay of stop squark where both chargino and sleptons are lighter 
   than the top squark. However in \cite{Sirunyan:2299392}  with 36 $fb^{-1}$ luminosity data the CMS collaboration put limit 
  on chargino mass in the compressed region where the chargino is produced in the strong production via top squark and decays to a $W$ boson and the lightest neutralino. The limit is displayed in magenta line.}
\label{fig:chargino_limit-plot}
\end{figure}

Another important aspect of the present SUSY spectrum is that it is capable of providing strong limit on electroweakinos owing to to strong production of top squarks which decay solely to charginos. In figure~\ref{fig:chargino_limit-plot} we depict the exclusion limit in chargino-neutralino plane at 13 TeV LHC with different integrated luminosities. We present the exclusion limits with $95 \%$ CLs corresponding to 36 $fb^{-1}$ luminosity by the purple line, whereas the limit corresponding to 300 $fb^{-1}$ luminosity is shown by the green line. The red line depicts the current CMS search limit~\cite{Sirunyan:2284431,Sirunyan:2299392} at 36 $fb^{-1}$ which is obtained from  tri-lepton + $\met$ final state originated from the decay of $\chi_1^\pm\,\chi_2^0$.  There is no experimental search in the dileptonic decay of stop squark where both chargino and sleptons are lighter than the top squark. However, in \cite{Sirunyan:2299392}  with 36 $fb^{-1}$ luminosity data the CMS collaboration put limit on chargino mass in the compressed region, represented by magenta line, where the chargino is produced in the strong production via top squark and decays to $W$ boson and the lightest neutralino. Evidently, our results compliment the current experimental exclusion limits in the regions where there are no existing bounds (very weak bounds). However, one should keep in mind that in the CMS study the electroweakinos are produced via weak interaction. Note that the CMS line shown in figure~\ref{fig:chargino_limit-plot} is also after considering the similar choice of mass gap $M_{\tilde{\ell}} = \alpha M_{\chi_1^+} + (1-\alpha) M_{\chi_1^0}$ with $\alpha = 0.5$ as we have pointed in our benchmark.
In the same reference~\cite{Sirunyan:2284431} it is also demonstrated with other extreme mass gap possibilities like $\alpha = 0.95, 0.05$ make the reach extremely poor due to soft leptons. In this study we can exclude the chargino mass up to 680 GeV for neutralino mass of 660 GeV for 13 TeV with 36 $fb^{-1}$ of integrated luminosity. Similarly, as evident from our model assumptions, the slepton mass up to 670 GeV can also be excluded for the same data.

Finally, we have tabulated all the benchmark points we have analyzed in table~\ref{tab:benchmark}. As we have 
already stated we vary the mass difference $\dmone$ and $\dmtwo$ in a controlled way to scan the whole parameter 
space. For all the benchmark points we have estimated LHC exclusion limit at 13TeV with integrated luminosity 
of 36 $\rm{fb}^{-1}$ and 300 $\rm{fb}^{-1}$. Also we have analyzed the discovery limit with integrated luminosity 
of 300 $\rm{fb}^{-1}$.

In passing by, we would like to mention that our variables can be useful in dedicated electroweakinos searches~\cite{Sirunyan:2299392} as well. Since signal and background topology for those searches are different, so one to one correspondence is not possible. However, keeping the compressed kinematics with equal velocity approximation in mind, we can still emphasize that part of the observable discussed in our work, if not all, will be useful to disentangle the background and signal events efficiently for those searches.


\section{Summary and Conclusion}\label{summary}
Presence of a light top squark ($M_{\tilde t} < 1$ TeV) is crucial for the naturalness of the Standard Model. 
Search for such a top squark at the LHC is very important and several studies are done by the CMS and ATLAS 
collaboration to put limit on the mass of the top squark which is very close to TeV scale. However, if the 
SUSY spectrum is compressed, the limit can go down significantly. Several experimental as well as theoretical 
efforts poured into searching such compressed top squark recently. Most of the studies for compressed top squark assume a simplified 
spectrum where top squark is the NLSP and the lightest neutralino being the LSP. There can be other possibilities where one or more 
intermediate states can be present in between such top squark and LSP. This may change the search strategies at the LHC and all these 
spectra need to be explored to carry out an exhaustive search for the top squark. 

One such possibility is having chargino and slepton in between top squark and the LSP while keeping the mass spectrum compressed.
In this case, top squark decay dominantly to a dileptonic final state along with missing energy from invisible LSPs and the SM neutrinos. Since the spectrum is compressed, 
all the final state particles are soft and it is very difficult to detect them. On the other hand  the dominant 
background is $t\bar t$ which is produced abundantly at the LHC owing to its huge production cross-section. 
Hence it is very important to minimize the background events efficiently in order to achieve a stringent limit 
on the top squark mass in the dileptonic scenario. Kinematics of the signal and the background processes can be 
used to separate them and in this paper we have studied potency of several kinematic variables which can be used  
to minimize the background. We notice that these variables encode the information about mass gaps between the 
top squark-chargino and slepton-neutralino pair and they are capable of providing a stringent limit on the top squark mass. 
We observe that depending on the mass gaps $\dmone$ and $\dmtwo$ a top squark with mass from 640 GeV to 715 GeV can be excluded at 2$\sigma$ with an integrated luminosity of 36 $\rm{fb}^{-1}$ data at the 13 TeV LHC. Similarly, the exclusion limit for the chargino achieved from this analysis is 680 for neutralino mass of 660 GeV and comparable limit for slepton (sneutrino) can also be obtained. The exclusion limit goes up to 840 GeV for 300  $\rm{fb}^{-1}$ data and at that luminosity it is possible to discover a top squark at 5$\sigma$ with mass up to 725 GeV.

\section{Acknowledgements}\label{sec:Acknowledgements}
This work was partially supported by funding available from the Department of Atomic Energy, Government of India, for the Regional Centre for Accelerator-based Particle Physics (RECAPP), Harish-Chandra Research Institute;  TDP project at Physical Research Laboratory, Department of Space, Government of India.  AKS acknowledges the support from Department of Science and Technology, Government of India under the fellowship reference number PDF/2017/002935 (SERB National Post Doctoral Fellowship). AKS also thank the Indian Association for the Cultivation of Science (IACS), Kolkata for administrative support.

\appendix 
\section{Benchmark Points}
 \begin{table}[htb!]
 	\centering
 	\begin{tabular}{||c|c|c||} \hline\hline
 		Parameters  & BP1 & BP2 \\ \hline
 		$\Delta M_{\tilde{t}\chi_1^+}$; $\Delta M_{\chi_1^+\tilde{\ell}} \approx \Delta M_{\tilde{\ell}\chi_1^0}$ & 30; 18  & 50; 45\\
 		\hline\hline 
 		$M_1$; $M_2$ & 500; 515 & 500; 568  \\    
 		$M_3$; $A_t$; $\mu$; $\tan\beta$  & \multicolumn{2}{c||}{ 3000; -3000; -2000; 20}  \\
 		\hline\hline
 		$m_{\widetilde g}$; $m_{\widetilde q_{L}}$; $m_{\widetilde q_{R}}$  &  \multicolumn{2}{c||}{2882; 2166; 2162} \\ 
 		$m_{\widetilde t_2}$; $m_{\widetilde b_1}$; $m_{\widetilde b_2}$ & \multicolumn{2}{c||}{2159; 2129; 2181}  \\
 		$m_{\widetilde\ell_{R}}$; $m_{\widetilde \tau_1}$;  $m_{\widetilde \tau_2}$ & \multicolumn{2}{c||}{2205; 2188; 2197}  \\
 		\hline\hline
 		$m_{\widetilde t_1}$ & 570.3 & 645.2  \\
 		$m_{\widetilde\chi^{\pm}_1}$ & 541.5 & 595.6  \\
 		$m_{\widetilde \ell_{L}}$; $m_{\widetilde\nu_{L}}$ & 519.3; 513.3 & 548.0; 542.1  \\
 		$m_{\widetilde\chi^0_1}$ & 501.9 & 502.0  \\
 		\hline
 		\hline
 		$m_h$ & 123.9 & 123.9  \\
 		$m_A$ & 1015.0     & 1013.0       \\
 		$\Omega h^2$ & 0.1102 & 0.1144 \\\hline
 		BR($b\to s\gamma$) &\multicolumn{2}{c||}{$3.57\times 10^{-4}$}  \\
 		BR($B_s\to \mu\mu$)&\multicolumn{2}{c||}{$2.77\times 10^{-9}$}  \\
 		\hline\hline
 	\end{tabular}
 	\caption{Low scale input parameters and the relevant sparticle masses for two of the chosen 
 		benchmark points satisfying all the collider, DM and low energy constraints. All the masses 
 		and dimensionful parameters are written in GeV units.}
 	\label{tab:bp_comp}
 \end{table}
In this appendix, we display two benchmark points with low scale input parameters and the relevant sparticle masses for the simplified model in our study. In both of these cases the neutralino is bino type and the lightest chargino is wino type. As shown in the table, the correct relic density can be obtained using  A-funnel region where $m_A\sim 2 m_{\chi_1^0}$. With a tuned $\tan\!\beta$ and trilinear coupling Higgs mass constraint is satisfied. We have fixed the gaugino mass parameter $M_1$ and varied $M_2$ and other soft SUSY breaking masses to achieve the desired mass spectrum as shown in the table. We have also satisfied the low energy flavor physics constraints, $b\to s\gamma$ and $B_s\to \mu\mu$ within $2\sigma$ limit.
 

\providecommand{\href}[2]{#2}
\addcontentsline{toc}{section}{References}
\bibliographystyle{JHEP}
\bibliography{bibliography}

\end{document}